\documentclass[12pt,preprint]{aastex}

\usepackage{graphicx}
\usepackage{psfig}
\usepackage{natbib}
\slugcomment{draft version}
\RequirePackage{longtable}%

\shorttitle{Phase-resolved  spectrum of 4U 1901+03}
\shortauthors{Y.J. Lei et al.}

\begin{document}

\title{Phase-resolved spectral analysis of 4U 1901+03 during its outburst}

\author{Ya-Juan Lei\altaffilmark{1}, Wei Chen\altaffilmark{1}, Jin-Lu Qu\altaffilmark{1}, Li-Ming Song\altaffilmark{1}, Shu Zhang\altaffilmark{1}, Yu Lu\altaffilmark{1}, Hao-Tong Zhang\altaffilmark{2}, Ti-Pei Li\altaffilmark{1,3}}

\affil{$^1$Particle Astrophysics Center, Institute of High Energy
Physics, Chinese Academy of Sciences, Beijing 100049, P.R. China;
leiyj@mail.ihep.ac.cn} \affil{$^2$National Astronomical
Observatories, Chinese Academy of Sciences, Beijing 100012, P.R.
China} \affil{$^3$Center for Astrophysics,Tsinghua University,
Beijing 100084, P.R. China}

\begin{abstract}

The high mass X-ray binary 4U 1901+03 was reported to have the pulse
profile evolving with the X-ray luminosity and energy during its
outburst in February-July 2003: the pulse peak changed from double
to single along with the decreasing luminosity. We have carried out
a detailed analysis on the contemporary phase-resolved energy
spectrum of 4U 1901+03 as observed by Rossi X-ray Timing Explorer
({\it RXTE}). We find that, the spectra are phase dependent. At the
begin of the outburst, the maximum of the optical depth for Compton
scattering is near the major phase peak. During the decay of the
outburst, the optical depth has the maximum being away from the main
peak of the pulse profile. For each observation, Fe K$\alpha$
emission line is detected in the phase-resolved spectra, and its
flux is constant across the pulse phases. This suggests an origin of
Fe emission from the accretion disk but not the surface of the
neutron star.

\end{abstract}


\section{Introduction}

Most of the known X-ray binary pulsars are the so-called high mass
X-ray binary (HMXB). They are usually the Be/X-ray binaries characterized by
 the transient nature.   The Be star is an early-type nonsupergiant star
with observable emission lines from the material in its
circumstellar disk (see Slettebak 1988 for a review). Be/X-ray
binaries usually show up with two types of outburst behavior: normal
outburst with low X-ray luminosity lasting for days-weeks, and giant
outbursts with higher X-ray luminosity ($L_{\rm X} \gtrsim 10^{37}$
erg s$^{-1}$) occurring irregularly in every several years. Giant
outbursts are thought to be driven by a dramatic expansion of the
disk surrounding the Be star, which leads to the formation of an
accretion disk around the compact object. Accompanying with the
giant outburst, the neutron star is detected with the pulsed
emission. The spectra of Be/X-ray binary are usually represented by
a cutoff power law shape (e.g., Coburn et al. 2002; Corbet et al.
2009; Crawford et al. 2009). The iron K feature between 6 and 7 keV
and low-energy absorption due to the cool material are as well
observable (White et al. 1983; Wilson et al. 2008).

%4U 1901+03 is a HMXB with a Be-type companion star.
4U 1901+03 was detected as  a Be/X-ray binary pulsar, one member of
the HMXB (Liu et al. 2006). The orbital period and eccentricity of
the system are measured as 22.58 d and 0.035, respectively. The
period of X-ray pulsation is about 2.73 s (Galloway et al. 2005).
The location of the source is R.A.=$19^{h}04^{m}13^{s}.4$ and
Dec=$+3^{o}09^{'}26^{''}$ (J2000.0) obtained by the observation with
{\it Uhuru} (Forman et al. 1978; Priedhorsky \& Terrell 1984). With
the observation of {\it RXTE} Proportional Counter Array (PCA),
Galloway et al. (2003a) obtained precise coordinates of
R.A.=$19^{h}03^{m}37^{s}.1$, Dec=$+3^{o}11^{'}31^{''}$.
%Cross-scan
%observations with {\it RXTE} Proportional Counter Array (PCA) and
%Uhuru located the source at R.A.=$19^{h}03^{m}37^{s}.1$,
%Dec=$+3^{o}11^{'}31^{''}$ (Galloway et al. 2003a).

There are two giant outbursts in the history of observations of 4U
1901+03. The first one was detected with Uhuru and Vela 5B in
1970-1971 (Forman et al. 1976; Priedhorsky \& Terrell 1984). The
second giant outburst took place in 2003 February, and was first
detected by All Sky Monitor (ASM) on {\it RXTE}, followed by series
of pointed {\it RXTE} observations over the next five months
(Galloway et al. 2003b). At the hard X-rays, contemporary
observations from INTEGRAL satellite are also available. During the
second giant outburst, the X-ray flux of the source reached a value
of $F_{2.5-25 keV}$$\sim$8$\times$$10^{-9}$ ergs cm$^{-2}$ s$^{-1}$.

Based on the RXTE observations, Galloway et al. (2005) made a
thorough study on the orbital parameters, preliminary X-ray spectral
analysis and pulse profiles from several observations for 4U 1901+03
during the 2003 giant outburst. Chen et al. (2008) carried out the
detailed analysis of the energy dependence of the pulse profile
along the outburst, and found that the pulse profile is correlated
with both the X-ray luminosity and photon energy (Wang \& Welter
1981; White et al. 1983; Nagase 1989; Mukerjee et al. 2000). The
phase-resolved spectral analysis is need for studying the emission
configure of Be/X-ray binary pulsar. Here we present for the first
time phase-resolved spectra of 4U 1901+03 using all available RXTE
data taken during the 2003 giant outburst. The paper is organized as
the follows: description of observations and spectral model in
Sec.2, results  in Sect.3 and discussion finally in Sec.4.

\section{Observation and data reduction}

The observations analyzed in this paper are from PCA and High Energy
X-Ray Timing Experiment (HEXTE) on board the {\it RXTE} satellite
from February 10 to July 16 2003. The PCA consists of 5 non-imaging,
coaligned Xe multiwire proportional counter units (PCUs) covering a
nominal energy rang from 2 to 60 keV. Only PCU0 and PCU2 data are
adopted in this work, which were on all the time during the
observation of 4U 1901+03. The HEXTE instrument consists of two
independent clusters (cluster A and cluster B) covering an energy
range from 15 to 250 keV. Because detector 2 of cluster B of HEXTE
lost its spectral capability and automatic gain control, only
cluster A data are analyzed in the work. We extract light curves and
spectra of PCA and HEXTE from intervals when the source has the
offset angle of less than $0.02^{o}$ and the limb of the earth is
more than $10^{o}$ with respect to the source direction. All HEXTE
data products are dead-time corrected using the HEASOFT
$\texttt{ftool hxtdead}$.

\begin{figure*}
\includegraphics[width=8cm,angle=270,clip]{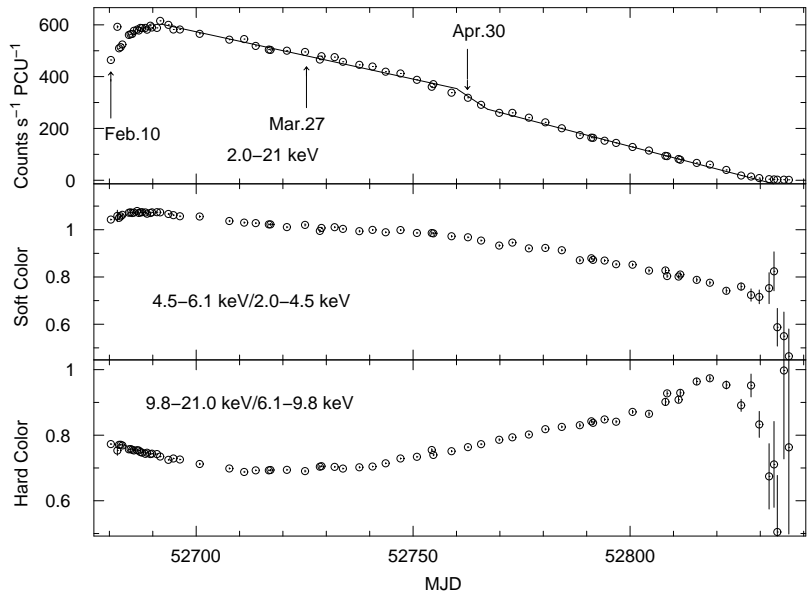}
\caption{ From top to bottom: the total light curve, the soft color,
and the hard color, respectively. } \label{fig1}
%\end{center}
\end{figure*}

The data of Standard-2 and GoodXenon modes of the PCA and Archive
and Science Event modes of the HEXTE are used to perform the spectra
and timing analysis. The total lightcurve is extracted with
Standard-2 mode data, and pulse-averaged X-ray spectra are extracted
with Standard-2 and Science Event mode data. The pulse profiles of
the chosen observations are extracted with GoodXenon, and the
phase-resolved spectra are extracted with GoodXenon and Science
Event data modes, using the software $\texttt{fasebin}$.
Figure~\ref{fig1} shows the lightcurve of 2.0-21.0 keV, the soft and
hard colors that are defined as the count rate ratios 4.5-6.1
keV/2.0-4.5 keV and 9.8-21.0 keV/6.1-9.8 keV, respectively.

The PCA background subtraction is carried out using the latest
versions of the appropriate background models, and a 1 \% systematic
error is added to the spectra to account for the calibration
uncertainties. Events in energy range $\sim$ 2.5-20 keV of the PCA
and 17-80 keV of the HEXTE (17-50 keV for the late observations of
the outburst) are selected for the spectral analysis with the
software XSPEC version 12.3.0p (Arnaud 1996; Dorman \& Arnaud 2001).

\subsection{Spectral Model}

\begin{figure*}
\includegraphics[width=10cm,angle=270,clip]{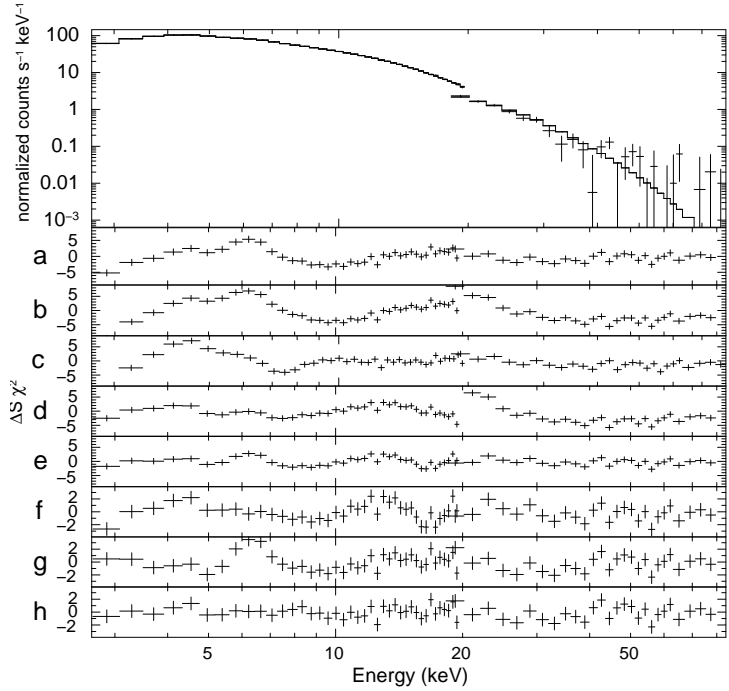}
\caption{ PCA spectrum for the observation of April 30 and the
best-fit model. The residuals are shown sequentially taking (a) the
cutoffpl model ($\chi^{2}_{\rm red}$=3.96), (b) the powerlaw+bbody
model ($\chi^{2}_{\rm red}$=13.15), (c) the powerlaw+bbody+gaussian
($\chi^{2}_{\rm red}$=6.38), (d) the bknpower model ($\chi^{2}_{\rm
red}$=9.04), (e) the bknpower+bbody model ($\chi^{2}_{\rm
red}$=2.09), (f) the bknpower+bbody+gaussian ($\chi^{2}_{\rm
red}$=1.75), (g) the compTT model ($\chi^{2}_{\rm red}$=1.69), (h)
the compTT+gaussian model ($\chi^{2}_{\rm red}$=0.98).} \label{fig2}
%\end{center}
\end{figure*}

According to the characteristic of the lightcurve (Fig. 1), we study
the phase-averaged spectra of ten typical observations, carried out
on February 10, 15, 22 and 23, March 27 and 30, April 30, May 25,
June 14 and 28, respectively, , covering the outburst profile of the
beginning, the peaking, the stepping down at the middle of decaying
and the ending tail. Various spectral models are used to fit the
phase-averaged spectra of these observations, e.g., cutoffpl,
powerlaw, bknpower, compTT, and a combination of each with a
blackbody component.
%Fits with the models of compTT is acceptable given with the rather small
% $\chi^{2}{\rm red}$.
A gaussian component centered at 6-7 keV which represents
fluorescent Fe line emission shows up as well in all spectra. A
model consisting of compTT with a spherical geometry (Titarchuk
1994) and a Gaussian component is statistically acceptable
($\chi^{2}_{\rm red}$ $\sim 1$) to fit the spectra (also see
Galloway et al. 2005). Due to low effective area PCA below 3 keV,
the column density $n_{\rm H}$ of neutral absorption can not be
constrained well. Therefore, for all spectral fits we fix $n_{\rm
H}$ at $1.2\times10^{22}$ cm$^{-2}$ (Galloway et al. 2005). Such a
model can well fit the data till April 30. For the observations near
the end of the outburst (May 25, June 14 and 28), an additional
component of a blackbody has to be introduced in order to have a
reasonable $\chi^{2}_{\rm red}$. The temperature of the blackbody is
$kT_{\rm bb}\sim$ 1 keV. As an example, Figure~\ref{fig2} shows the
fitting results with various models to the data of April 30.
%where the model of CompTT plus a Gaussian component is the best choice.

\begin{figure*}
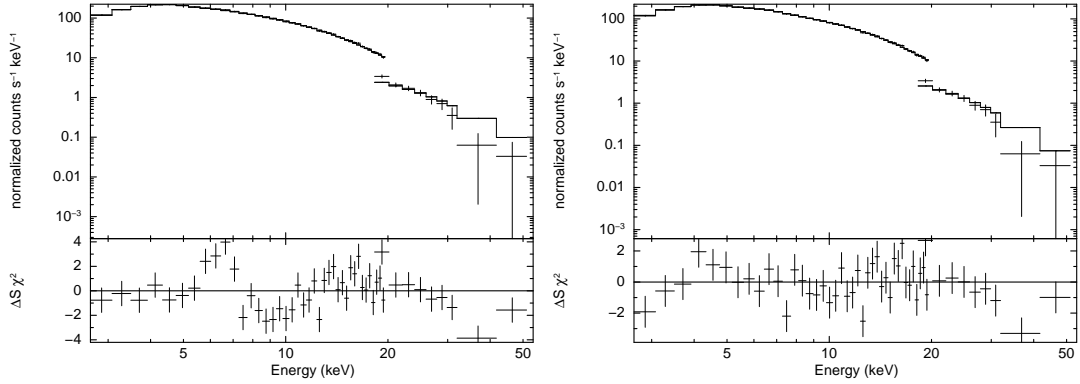

\begin{center}
\includegraphics[width=5cm,angle=270,clip]{f3a.ps}
\includegraphics[width=5cm,angle=270,clip]{f3b.ps}
\caption{The fitting results of the phase-resolved spectrum of the
phase 5 of the observation April 30, the left panel shows the
fitting result by the model without Gaussian component, the residual
shows obvious Fe line profile; the right panel shows the fitting
result with the model containing the Gaussian component, the
residual is removed.} \label{fig3}
\end{center}
\end{figure*}

We choose three typical observations on February 10, March 27, and
April 30 to analyze the evolvement of the phase-resolved spectra.
The phase-resolved spectra are extracted with eight phase bins.
% by GoodXenon mode data of the PCA and Science Event mode data of the HEXTE.
% The spectrum for each phase bin consists of both the pulsed
% and non-pulsed emission, and we regard it as the 'continuum'. The
% phase-resolved spectra of the pulsed emission are obtained by
% subtracting the phase bin with the lowest intensity regarded as the
% non-pulsed part.
Figure~\ref{fig3} shows the spectra from the data of April 30, for
phase bin No.5 (0.4375-0.5625). The fitting results of the spectra
show the gaussian component is needed in the spectrum.

\section{Results}

\subsection{Variation of soft and hard colors}

Figure~\ref{fig1} shows a stepping down feature of  the light curve,
happened at time around  April 30, during the decay of the outburst.
The diagnosis on the pulse profile revealed a change from double
peaks to single peak at this time (Chen et al. 2008). Accompanying
to this as well are modifications in the fit model of the spectrum
(Galloway et al. 2005; Chen et al. 2008). Along the decay of the
outburst, the hard and the soft colors have the different
evolutions: the soft color evolves similar to the light curve, while
the hard color shows the opposite and ends up with a rapid decrease.
The soft color and the hard color suggest an overall trend that the
spectrum softens at the begin, but hardens with the decay of the
outburst, and turns to soft at the end.

\subsection{Phase-averaged spectra}

Table 1 shows the  results of the fit models on ten typical
observations described in Sect.2, i.e., the beginning, the peaking,
near stepping down and near the ending. The results show the
temperature of the seed photons ($T_{\rm 0}$) in the Comptonization
model is about 1 keV, and decreases slightly during the decay of the
outburst. The temperature of the hot electronic population (kT)
remained at about 5 keV. The optical depths of the observations from
February 10 to April 30 are consistent with a flat distribution
within the error bars, and an average is derived as $\sim$ 5. At the
end of the outburst, the optical depths are obtained as 7-8. We
notice that, the Fe K$\alpha$ line emission with equal width $\sim$
100 eV presents at around 6.5 keV in the energy spectrum, and its
flux varies between $\sim$1-10 $\times$ $10^{-11}$ ergs cm$^{-2}$
s$^{-1}$, showing a trend of increasing with the intensity. At the
end of outburst, an additional spectral component of a blackbody is
needed for spectral fittings.

\subsection{Phase-resolved spectra}

\begin{figure*}

\includegraphics[width=14cm,angle=0,clip]{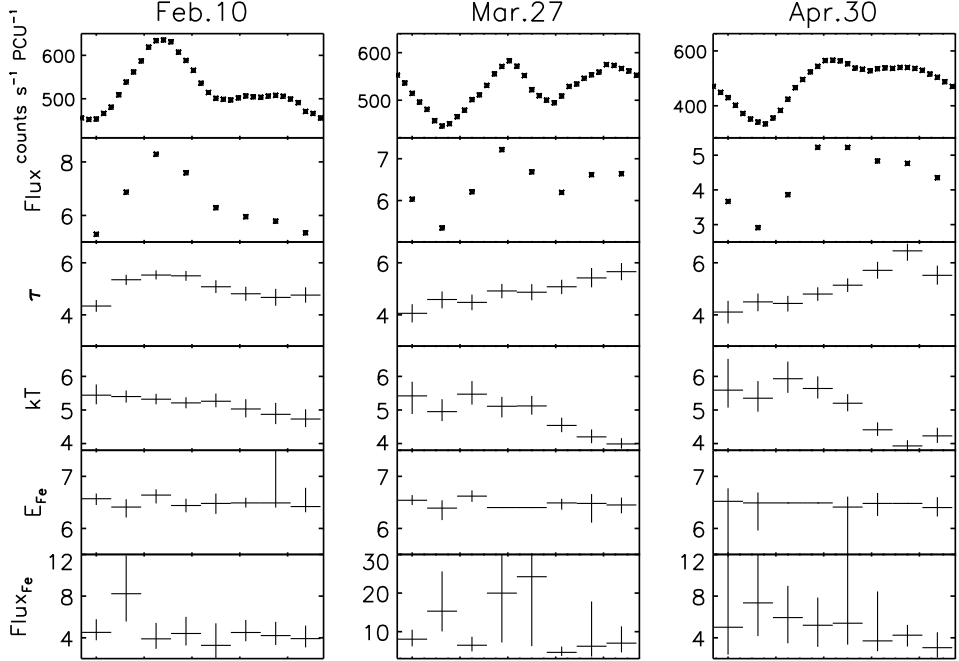}
\begin{center}
\caption{The phase-resolved spectral parameters as a function of
pulse phases. From top to bottom: the pulse profiles of 32 phase
bins obtained from all PCA channels (2-60 keV), the flux of 3-30 keV
of 8 phase bins in units of $10^{-9}$ ergs cm$^{-2}$ s$^{-1}$,
optical depth for Compton scattering, the temperature of the
scattering electron cloud, the central energy and flux of Fe
emission line. The central energies of Fe line are fixed for phases
4 and 5 of March 27 and phases 3 and 4 of April 30.} \label{fig4}
\end{center}
\end{figure*}

\begin{figure*}
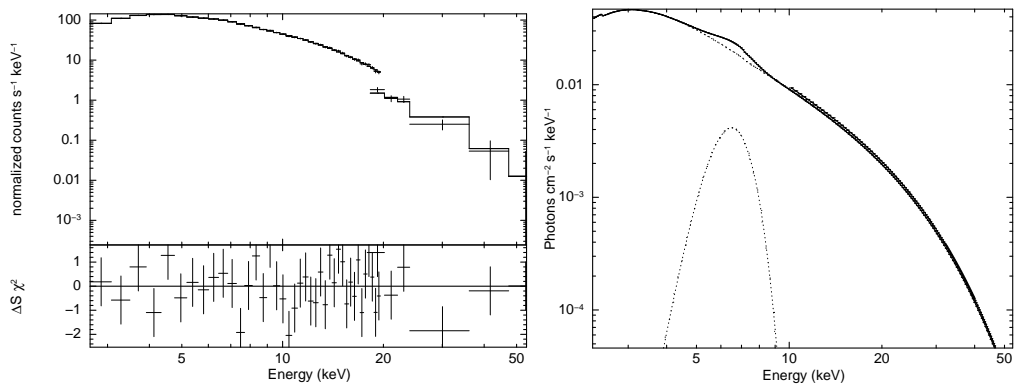

\begin{center}
\includegraphics[width=5cm,angle=270,clip]{f5a.ps}
\includegraphics[width=5cm,angle=270,clip]{f5b.ps}
\caption{The fitting result of the spectrum of the phase 2 of the
observation April 30. The left panel shows the fitting result and
residual, the right panel shows the model components of
(compTT+gaus). } \label{fig5}
\end{center}
\end{figure*}

For studying the evolution  of the phase-resolved spectra, we choose
the typical  observations with the  high signal-to-noise ratio,
i.e., on February 10, March 27 and April 30. The spin lightcurve of
each observation is subdivided into 8 phases, for each the energy
spectrum is extracted, and the flux is shown in Figure~\ref{fig4}.
Figure~\ref{fig4} shows the phase dependence of the spectra of the
three observations. The pulse profile of the three observations are
best presented in Figure 4, each with 32 phase bins. At the
beginning of the outburst (February 10), the optical depth maximizes
near the major phase peak. However, during the decay of the
outburst, such maximum moves towards  the second one, and the
temperature of the scattering electron has a trend of
anti-correlation with the optical depth. The center energy of Fe
line is almost around 6.5 keV. For each observation, the fluxes of
iron line are likely constant across the whole pulse phases.  The
spectra from the phases (1 for February 10, 2 for March 27 and 2 for
April 30) of the lowest intensity present as well the existence of
the obvious Fe feature, as shown in Figure~\ref{fig5} is an example
from the phase 2 of the data of April 30, with the fluxes comparable
to those from what else phases (Fig.~\ref{fig4}).

\subsection{Spectral Ratios}

\begin{figure*}
\includegraphics[width=10cm,angle=270,clip]{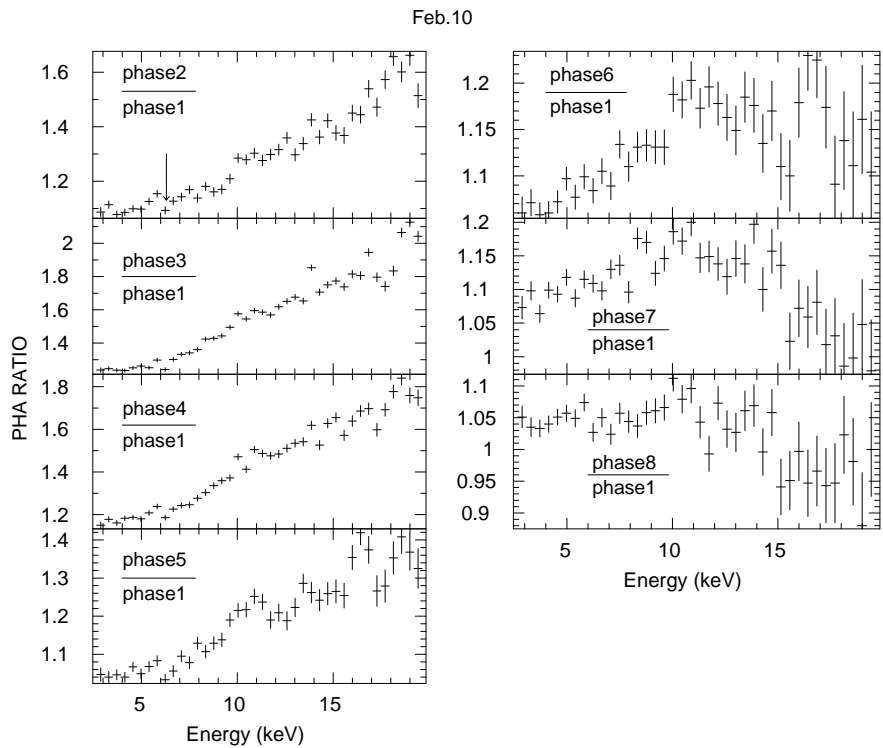}
\caption{ Ratios of intensity in narrow energy bands from February
10 phase 2-8 spectra to intensity from phase 1 spectrum. }
\label{fig6}
%\end{center}
\end{figure*}

\begin{figure*}
\includegraphics[width=10cm,angle=270,clip]{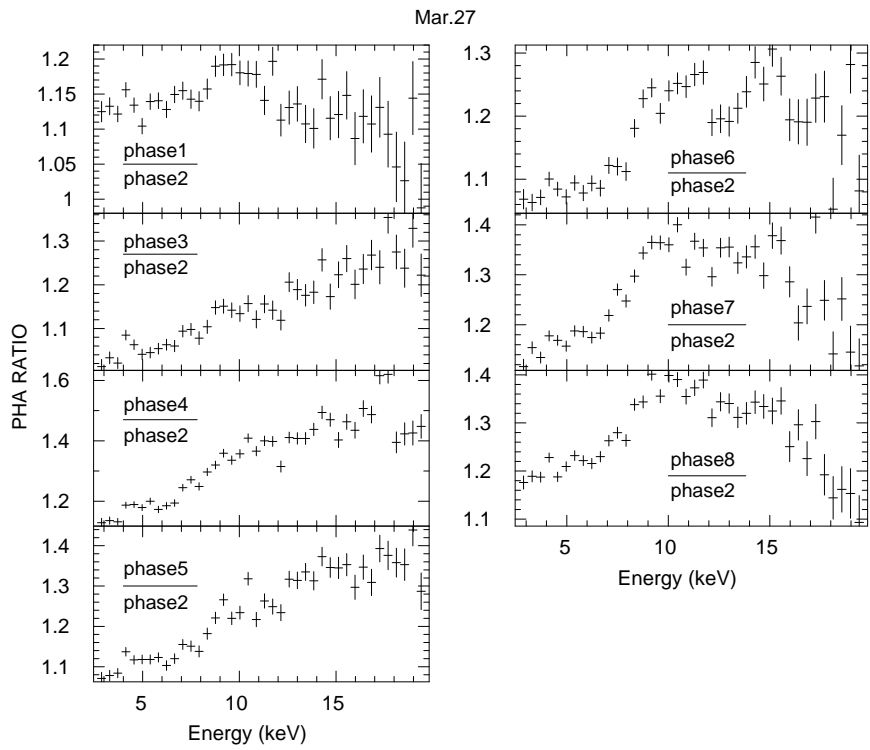}
\caption{Similar to Figure 6, but for March 27. } \label{fig7}
%\end{center}
\end{figure*}

\begin{figure*}
\includegraphics[width=10cm,angle=270,clip]{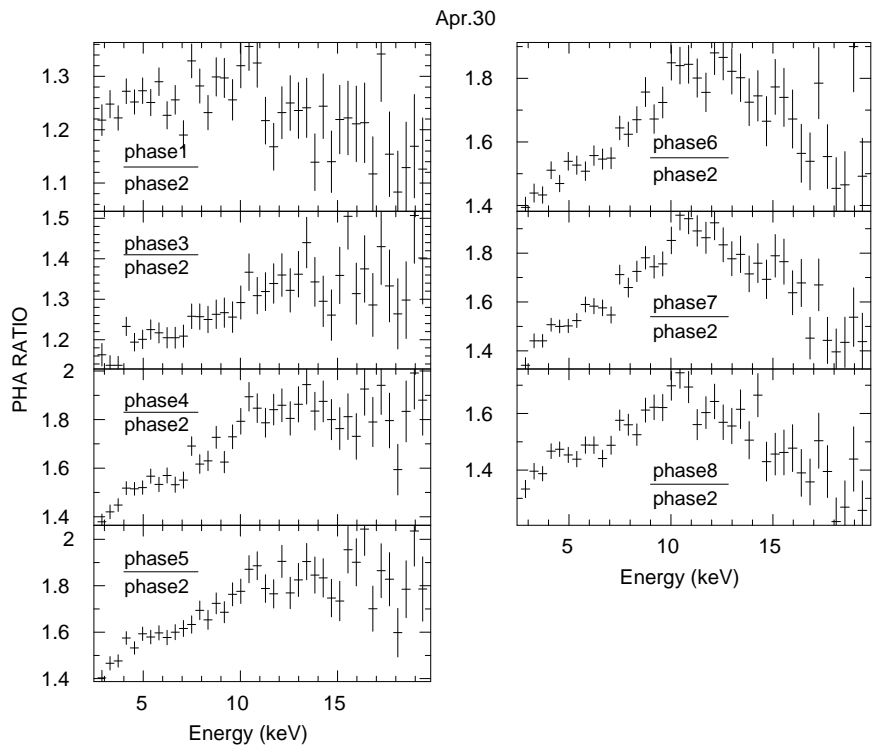}
\caption{Similar to Figure 6, but for April 30. } \label{fig8}
%\end{center}
\end{figure*}

The detailed phase-to-phase variation is best illustrated by ratios
of the phase-resolved spectra to the spectrum of the phase with
minimum count rate (Leahy \& Matsuoka 1990). We show the
phase-resolved spectral ratios of the three observations. The
minimum-count-rate spectra are from phase 1 for observation of
February 10, phase 2 for both observations of March 27 and April 30.
The phase ratios are then presented separately in
Figure.~\ref{fig6}-\ref{fig8}. There are dips in the PHA plots at
$\sim$ 6.5 keV (as marked by the arrow in the first panel of
Fig.~\ref{fig6}). The dip is due to that flux of Fe emission line
that only occurs in non-pulse component, this is consistent with the
results of the phase-resolved spectra.

For the phase 2 of February 10, at the beginning of the main pulse,
 the ratios increase with energy. The ratio increases with energies rapidly
at the phase 3 where the main pulse is peaking, but slowly
 at the phases (4 and 5) beyond. As is the case as well for
at  the end of the main pulse (phase 6), till the energies around 10
keV, beyond which the ratio remains almost constant. Such a trend
holds more or less in phases 7 and 8, corresponding to the second
pulse, but with a turn over of the ratio at energies larger than 10
keV. For March 27, the properties of PHA ratios are similar to those
for February 10. For April 30, phases 3, 4 and 5 correspond to the
main pulse peak. The PHA ratios of these phases are increase with
energy, but are constant at above 10 keV. Phases 6, 7, 8 and 1 are
around the second pulse, and their  PHA ratios increase at low
energy ($<$10 keV), and decrease at 10-20 keV. The overall spectral
ratio of April 30 is different from those of February 10 and March
27.

The PHA ratios of the phase-resolved to the phase with minimum count
rate are increasing with energy, illuminating that the spectra of
the pulsed are generally harder. The stronger the flux of the
pulsed, the harder the spectrum. At the phases around the second
pulse, the spectrum decreases abruptly at higher energies.

\section{Discussion}

The X-ray radiation modes of Be/X-ray binary pulsar is generally
thought to have tight relation with the luminosity (Parmar et al.
1989). At high luminosity ($\gtrsim$ 10$^{37}$ ergs s$^{-1}$), the
accretion flow onto the magnetic pole can be decelerated via the
radiative shock formed near the neutron star surface (Wang \& Frank
1981). The emitting plasma will be compressed under the shocked
region, where the photons can only escape from the sides of the
column (fan-beam mode). Under some circumstances, a pencil-beam may
still emerge (Nagel 1981). At lower luminosity ($\lesssim$ 10$^{37}$
ergs s$^{-1}$) the infalling material may be decelerated in a
collisionless shock above the neutron star surface (Basko \& Sunyaev
1975; Kirk \& Galloway 1981). In such a case,  a thin emitting
region or effects of the strong magnetic field can cause a
pencil-beam of emission to be formed (e.g., M\'{e}sz\'{a}ros et al.
1983). The various models could be verified by the observational
properties of X-ray pulsars.

We have analyzed the phase-resolved spectra of 4U 1901+03, and found
that the optical depth and the temperature of the scattering
electron are related to the pulse phases, which is a common feature
of X-ray pulse binary (e.g., La Barbera et al. 2003).
%The optical depth of the pulsed is larger than that of the non-pulsed.
Spectral ratios indicate the main pulse peak has the hardest
spectrum, which is common property of accreting pulsars (Hickox \&
Vrtilek 2005; Tsygankov et al. 2007). Our results show, at the
beginning of the outburst (February 10) where the luminosity
($\gtrsim$ 10$^{37}$ ergs s$^{-1}$), the emission of the main pulse
has the possible origins of the fan-beam. The main pulse has the
large optical depth, could be due to that, during the main peak, the
angle between the column axis and the observer's line of sight has
the highest value so that the observer is looking almost along the
beam (Klochkov et al. 2008). During the decay of the outburst, our
results are consistent with the emission configuration that the main
peak from the fan-beam and the second peak from pencil-beam (Chen et
al. 2008), the main peak that has the hard spectrum is due to that
high energy photons are more liable to escape in a fan-beam from a
hot region close to the footstep and perpendicular to the accretion
column (Basko \& Sunyaev 1976; White et al. 1983; Klochkov et al.
2008), which is consistent with the low optical depth and high
temperature of the scattering electron of the main peak. The second
peak is produced by the low energy photons escaping in a pencil-beam
along the direction of the accretion column where the optical depth
is lager and the electron temperature is lower. These results are
accordant to those of Chen et al. (2008) who analyzed
energy-resolved pulse profiles in details. They concluded as well
that, the fan-beam contributes to the main pulse peak and
pencil-beam to the second pulse peak. In addition, in some energy
bands the flux varies by almost a factor of two between different
pulse phases (Chen et al. 2008), both the angles of the spin-axis
from the viewing direction and between the magnetic pole and the
spin-axis are substantial. Therefore, it is likely that emission
from both magnetic poles contribute to the pulse profile.

We notice that the evolution of hard color is similar to that of
pulse fraction (see Chen et al. 2008), and spectral ratios show that
the spectra of the pulsed are harder than those of the non-pulsed
under larger pulsed fluxes. This indicates that the high energy
photons contribute mostly to the pulsed emission, which is
consistent with the evolvement of the pulse fraction shown by Figure
4 of Chen et al. (2008). Perhaps the lower energy radiation is
dominated by photons from the accretion column walls that hit the
neutron star surface and are reprocessed, which would lead to a
lower pulse fraction.
%The electron
%temperatures for each phase are similar between the continuum and
%the pulsed. Since that the high energy photons are produced via
%Compotonizations of the hot electron population off the soft
%photons, a larger optical depth will lead to more productive of the
%high energy photons. Therefore, that the optical depths of the
%pulsed emission are larger is naturally expected.

Fe K$\alpha$ emission line with equivalent widths of several hundred
eV is usually detected in X-ray pulsar, may be caused by
illumination of neutral or partially ionized material in the
accretion disk, stellar wind of the high-mass companion, or material
in the line of sight or in the accretion column (Pravdo et al. 1977;
Basko 1980; Nagase 1985; Paul et al. 2002; Naik et al. 2005). Our
results show that prominent Fe emission with equivalent width $\sim$
100 eV is detected in the spectra of 4U 1901+03. The flux of Fe
emission line is increasing with the X-ray luminosity along the
outburst (Fig. 1 and Table 1), and is proportional to that of
continuum intensity and increases with increasing of the accretion
rate (Suchy et al. 2008). For each observation, Fe line is detected
in the spectra of the continuum in each pulse phase and its flux is
constant within errors, and is not correlated with pulse phases.
Therefore, the Fe line of 4U 1901+03 is not likely to origin from
the accretion column of magnetic polar cap. Furthermore, the
spectral fit of the phase-resolved spectrum shows that, the center
energy of Fe neither change among the pulse phases nor evolve along
the outburst of 4U 1901+03. We therefore believe that, during the
outburst of 4U 1901+03, Fe emissions should come from a region in
the accretion disk but not the surface of the neutron star.

\section*{Acknowledgments}
The authors are grateful to the referee for the helpful comments. We
are thankful for Prof. F.J. Lu for the useful discussions. This work
is subsidized by the Special Funds for Major State Basic Research
Projects and by the Natural Science Foundation of China for support
via NSFC 10903005, 10773017, 10325313, 10521001, 10733010, Xinjiang
Uygur Autonomous Region of China (Program 200821164) and program of
the Light in Chinese Western Region (LCWR) under grant LHXZ 200802,
the CAS key project via KJCX2-YWT03, and 973 Program 2009CB824800.

\begin{table*}
\begin{center}
\caption{\bf ~~Fit parameters for the spectra.}
\medskip
\begin{tabular}{l c c c c c c c c}
\hline
\hline
      date & $T_{\rm 0}$         & kT           & $\tau$   & $E_{\rm Fe}$ & Flux$_{\rm Fe}$       & $\chi^{2}_{\rm red}$\\
           &   (keV)             & (keV)            &          & (keV)      &     &        \\
\hline
  Feb.10   &$1.027_{-0.072}^{+0.026}$  & $5.28_{-0.27}^{+0.15}$ & $4.95_{-0.15}^{+0.45}$  &$6.45_{-0.58}^{+0.13}$   & $6.35_{-1.88}^{+14.13}$& 1.14     \\
  Feb.15   &$1.117_{-0.021}^{+0.019}$  & $5.09_{-0.07}^{+0.06}$ & $4.96_{-0.09}^{+0.09}$  &$6.52_{-0.12}^{+0.12}$   & $7.85_{-1.85}^{+2.46}$ &0.91        \\

  Feb.22   &$1.128_{-0.027}^{+0.023}$  & $4.87_{-0.09}^{+0.11}$ & $5.01_{-0.14}^{+0.15}$  &$6.54_{-0.12}^{+0.11}$   & $8.98_{-2.20}^{+3.04}$ &0.92      \\

  Feb.23   &$1.114_{-0.030}^{+0.025}$  & $4.86_{-0.09}^{+0.14}$ & $4.98_{-0.15}^{+0.17}$  &$6.53_{-0.11}^{+0.11}$   & $9.85_{-2.40}^{+3.65}$ & 0.74     \\
  Mar.27   &$1.028_{-0.063}^{+0.038}$  & $4.70_{-0.21}^{+0.21}$ & $4.98_{-0.25}^{+0.37}$  &$6.54_{-0.20}^{+0.14}$   & $9.13_{-3.01}^{+7.46}$ & 0.65     \\
  Mar.30   &$0.991_{-0.046}^{+0.029}$  & $4.70_{-0.13}^{+0.12}$ & $5.09_{-0.17}^{+0.24}$  &$6.46_{-0.17}^{+0.12}$   & $8.87_{-2.49}^{+5.54}$ & 0.92     \\
  Apr.30   &$0.898_{-0.049}^{+0.036}$  & $4.90_{-0.19}^{+0.21}$ & $5.15_{-0.22}^{+0.26}$  &$6.41_{-0.16}^{+0.12}$   & $5.27_{-1.57}^{+3.04}$ & 0.98     \\

\hline
\hline
date   &$T_{\rm 0}$          & kT                 & $\tau$   & $E_{\rm Fe}$ &Flux$_{\rm Fe}$    & $kT_{\rm bb}$  &$norm_{\rm bb}$& $\chi^{2}_{\rm red}$   \\
       &   (keV)             & (keV)              &        & (keV)      &    & (keV)    &     &   \\
\hline
May.25&$0.80_{-0.21}^{+0.11}$  &$4.61_{-0.23}^{+0.23}$&$7.38_{-0.97}^{+0.92}$&$6.48_{-0.16}^{+0.12}$&$1.53_{-0.41}^{+0.61}$ &$1.15_{-0.17}^{+0.08}$&$31_{-6}^{+37}$&0.56\\
Jun.14 &$1.00(fixed)       $   &$4.75_{-0.32}^{+0.32}$&$7.23_{-0.85}^{+1.32}$&$6.67_{-0.15}^{+0.14}$&$1.31_{-0.41}^{+0.58}$ &$0.92_{-0.05}^{+0.06}$&$52_{-7}^{+9}$&0.71\\
Jun.28 &$0.52_{-0.52}^{+1.03}$ &$5.01_{-0.33}^{+0.40}$&$7.67_{-0.98}^{+1.30}$&$6.59_{-0.18}^{+0.20}$&$0.54_{-0.22}^{+0.53}$ &$0.91_{-0.13}^{+0.06}$&$25_{-6}^{+16}$&0.80\\
\hline
\end{tabular}
\end{center}
\footnotesize \tablecomments{$T_{\rm 0}$ is the temperature of the
soft seed photons for the Comptonization, kT is the electron
temperature, $\tau$ is the optical depth of the scattering cloud in
a spherical geometry. $E_{\rm Fe}$ is the central energy of Gaussian
emission line, and Flux$_{\rm Fe}$ is the flux of Gaussian emission
line in units of $10^{-11}$ ergs cm$^{-2}$ s$^{-1}$. $kT_{\rm bb}$
is the temperature of blackbody, and $norm_{\rm bb}$ is the
normalization of blackbody.}
\end{table*}

\end{document}